\documentclass[conference]{IEEEtran}
\IEEEoverridecommandlockouts

\usepackage{booktabs}

\usepackage{cite}
\usepackage{amsmath,amssymb,amsfonts}
\usepackage{algorithmic}
\usepackage{graphicx}
\usepackage{textcomp}
\usepackage{xcolor}
\def\BibTeX{{\rm B\kern-.05em{\sc i\kern-.025em b}\kern-.08em
    T\kern-.1667em\lower.7ex\hbox{E}\kern-.125emX}}
\begin{document}

\title{EDGChain-E: A Decentralized Git-Based Framework for Versioning Encrypted Energy Data \\

}

\author{

    \IEEEauthorblockN{Alper Alimoglu}
    \IEEEauthorblockA{\textit{Huawei Turkey R\&D Center} \\
        \textit{Cloud Business DC Department} \\
        Istanbul, Turkey \\
        alper.alimoglu@huawei.com}
    \and
    \IEEEauthorblockN{Kamil Erdayandi}
    \IEEEauthorblockA{\textit{Computer Engineering Department } \\ \textit{ Faculty of Engineering} \\
        \textit{Marmara University}, Istanbul, Turkey \\
        kamil.erdayandi@marmara.edu.tr}
        \IEEEauthorblockA{\textit{Department of Computer Science } \\ \textit{The University of Manchester},  UK \\       kamil.erdayandi@manchester.ac.uk}
    \and
    \IEEEauthorblockN{Mustafa A. Mustafa}
    
        \IEEEauthorblockA{\textit{Department of Computer Science } \\ \textit{The University of Manchester},  UK \\       mustafa.mustafa@manchester.ac.uk}       
    \and
        \IEEEauthorblockN{Ümit Cali}
    \IEEEauthorblockA{\textit{School of Physics}\\ \textit{Engineering and Technology} \\
        \textit{University of York}, York, UK \\
        umit.cali@york.ac.uk}
}

\author{
    \IEEEauthorblockN{
        Alper Alimoglu\IEEEauthorrefmark{1},
        Kamil Erdayandi\IEEEauthorrefmark{2}\IEEEauthorrefmark{3},  
        Mustafa A. Mustafa\IEEEauthorrefmark{3}\IEEEauthorrefmark{4}, and
        Ümit Cali\IEEEauthorrefmark{5}}
\IEEEauthorblockA{
    \IEEEauthorrefmark{1}\textit{Huawei Turkey R\&D Center}, \textit{Cloud Business DC Department}, Turkey\\
    \IEEEauthorrefmark{2}\textit{Computer Engineering Department}, \textit{Marmara University}, Turkey\\
    \IEEEauthorrefmark{3}\textit{Department of Computer Science }, \textit{The University of Manchester}, UK\\
    \IEEEauthorrefmark{4}\textit{COSIC}, \textit{KU Leuven}, Belgium \\
    \IEEEauthorrefmark{5}\textit{School of Physics}, \textit{University of York}, UK\\
    Email: alper.alimoglu@huawei.com, kamil.erdayandi@marmara.edu.tr
    }
    
}

\maketitle

\begin{abstract}
This paper proposes a new decentralized framework, named EDGChain-E (Encrypted-Data-Git Chain for Energy),  designed to manage version-controlled, encrypted energy data using blockchain and the InterPlanetary File System. The framework incorporates a Decentralized Autonomous Organization (DAO) to orchestrate collaborative data governance across the lifecycle of energy research and operations, such as smart grid monitoring, demand forecasting, and peer-to-peer energy trading.    
In EDGChain-E, initial commits capture the full encrypted datasets—such as smart meter readings or grid telemetry—while subsequent updates are tracked as encrypted Git patches, ensuring integrity, traceability, and privacy. This versioning mechanism supports secure collaboration across multiple stakeholders (e.g., utilities, researchers, regulators) without compromising sensitive or regulated information.
We highlight the framework's capability to maintain FAIR-compliant (Findable, Accessible, Interoperable, Reusable) provenance of encrypted data. By embedding hash-based content identifiers in Merkle trees, the system enables transparent, auditable, and immutable tracking of data changes, thereby supporting reproducibility and trust in decentralized energy applications.
\end{abstract}


\section{Introduction}
Encrypted datasets are essential in scientific research for ensuring reliable, verifiable, and reproducible results in a secure and privacy-preserving way. To make such data both secure and usable, researchers are increasingly adopting the FAIR principles—Findable, Accessible, Interoperable, and Reusable~\cite{fair}. These principles guide the organization of data to ensure that data can be securely shared and reused without compromising sensitive information. By following FAIR standards, researchers could enable privacy-preserving collaboration and meet the rigorous requirements for reproducibility, especially when dealing with private or regulated data.

However, widely used platforms like GitHub, GitLab, and Software Heritage~\cite{heritage} are designed primarily for code versioning and lack native support for managing large, \textit{encrypted} datasets. This limitation is particularly evident in domains where datasets are both large and \textit{sensitive}. In the energy sector, for example, encrypted datasets are critical for maintaining privacy in applications like smart meter data analysis~\cite{asghar2017smart}, decentralized energy clearance~\cite{erdayandi2024pp} and billing~\cite{erdayandi2025privacy}, and grid monitoring~\cite{kumar2019smart}. When data is not protected, raw datasets often contain personally identifiable or commercially sensitive information, necessitating strict privacy-preserving~\cite{erdayandi2022towards} and security measures~\cite{cali2023digital}. 
Commonly used platforms such as GitHub, GitLab, and Software Heritage cannot guarantee compliance with FAIR or security requirements in energy domain considering the need for specialized infrastructure tailored to version-controlled, \textit{encrypted} energy data.

Repositories such as Zenodo~\cite{zenodo} introduce further challenges.
While they assign persistent DOIs to uploaded datasets, they prevent any further modifications and do not offer version tracking capabilities.
This rigid approach is incompatible with the dynamic nature of encrypted energy data, 
which may require frequent updates. 
Moreover, Zenodo lacks tools for incremental updates and audit trails. To address these issues, a secure framework should allow controlled workflows (e.g., decrypt-edit-re-encrypt), automatically record version history for sensitive data.
Such a system would preserve FAIR compliance while protecting sensitive information and maintaining data integrity.

\begin{table*}[t]
\centering
\caption{Comparison of frameworks for e-Science versioning and EDGChain-E proposal}
\label{tab:framework-comparison}
\begin{tabular}{@{}p{3cm}p{6.5cm}p{6.5cm}@{}}
\toprule
\textbf{Framework} & \textbf{Strengths} & \textbf{Limitations} \\
\midrule
GitHub / GitLab &
\begin{itemize}
  \item Distributed version control (Git)
  \item Rich collaboration features (issues, pull requests, CI/CD)
  \item Private repositories and team access
\end{itemize} &
\begin{itemize}
  \item No end-to-end encryption
  \item Limited binary/large data support (e.g., Git Large File Storage around 2 GB)
  \item Mutable history (e.g., rebasing, force-push)
  \item Poor metadata support for datasets
\end{itemize} \\
\addlinespace
Software Heritage &
\begin{itemize}
  \item Immutable public archive of software artifacts
  \item Persistent identifiers
  \item Tracks provenance across code repositories
\end{itemize} &
\begin{itemize}
  \item Read-only access (no updates/collaboration)
  \item No access control or encryption
  \item Limited to open-source code
  \item No support for large datasets
\end{itemize} \\
\addlinespace
Zenodo / Figshare &
\begin{itemize}
  \item Persistent DOIs for citation
  \item Rich metadata and license support
  \item Supports embargoed data during peer review
  \item Backed by leading research institutions (e.g., CERN)
\end{itemize} &
\begin{itemize}
  \item Snapshot-based versioning only
  \item Centralized trust model
  \item No encryption
  \item Limited support for large data ($>$50 GB)
\end{itemize} \\
\addlinespace
DVC &
\begin{itemize}
  \item Git-like versioning for large datasets
  \item Flexible storage backend (Amazon S3, Google Cloud Storage, Azure Blob Storage, etc.)
  \item ML pipeline and experiment tracking
\end{itemize} &
\begin{itemize}
  \item No access control or encryption
  \item Requires manual setup and Git proficiency
  \item No global provenance or history ledger
  \item Lacks persistent identifiers
\end{itemize} \\
\addlinespace
IPFS + Blockchain &
\begin{itemize}
  \item Content-addressed, integrity-checked storage
  \item Immutable record via blockchain
  \item Smart contracts for version tracking
\end{itemize} &
\begin{itemize}
  \item Encryption not native (needs external layer)
  \item High complexity for setup/integration
  \item IPFS not ideal for mutable or frequently updated data
  \item User interfaces are less mature
\end{itemize} \\
\addlinespace
EDGChain-E (Proposed) &
\begin{itemize}
  \item Native versioning and encryption
  \item Blockchain-backed immutability and audit trails
  \item Decentralized, trustless data/code sharing
  \item Metadata-rich, science-oriented design
\end{itemize} &
\begin{itemize}
  \item Requires adoption and tooling support
  \item Needs performance tuning for large-scale workloads
  \item Usability and UX still under development
  \item No access control
\end{itemize} \\
\bottomrule
\end{tabular}
\end{table*}

Furthermore, existing solutions often lack mechanisms to ensure data integrity and provenance over time. Without robust version control and audit trails, it becomes challenging to verify its evolution throughout the research life cycle. 
This gap underscores the necessity for a system that not only secures data but also provides transparent and {\textit{immutable}} records of changes, thereby enhancing trust and accountability in scientific research.

Efforts to overcome these challenges have led to the development of tools such as Data Version Control (DVC)~\cite{dvc}, often marketed as a Git-like system for data.
While DVC allows tracking of datasets via lightweight \texttt{.dvc} files, the raw data is stored externally in cloud or local storage systems, limiting collaborative workflows.
When applied to encrypted data file, DVC presents several limitations:
\begin{itemize}
   
    \item \textbf{No Version Integrity}: There is no cryptographic guarantee of the integrity of data versions. 
    \item \textbf{Need for Manual Work}: Updates require manual re-encryption and re-upload.
    \item \textbf{FAIR Non-Compliance}: Encrypted data often lacks standardized metadata, reducing findability and reusability.
\end{itemize}

A more robust solution would combine DVC-style versioning with encryption-aware data repositories that:
\begin{itemize}
   
    \item Embed cryptographic hashes into version logs for integrity,
    \item Generate FAIR-compliant metadata including update history and decryption protocols.
\end{itemize}

While platforms such as GitLab and Software Heritage provide mechanisms for archiving research software, they fall short of offering strong guarantees for long-term scientific reproducibility. 
GitLab, as a privately owned service, permits users to modify or delete repositories, which compromises the permanence and verifiability of research artifacts.
Software Heritage improves upon this by generating intrinsic, content-based identifiers; however, its reliance on centralized, publicly funded infrastructure raises concerns about long-term sustainability.
Decentralized alternatives, such as repositories that assign Digital Object Identifiers (DOIs) or systems built on the InterPlanetary File System (IPFS), provide a more resilient infrastructure for research reproducibility by enhancing data integrity and reducing dependence on centralized authorities. However, encryption is not native in this framework -- an external layer is needed.
Note that IPFS is a decentralized, peer-to-peer file storage protocol
that ensures content integrity through content-addressing.

To address existing limitations in  data storage and ownership tracking, we propose the EDGChain-E framework -- a novel architecture that leverages blockchain technology and decentralized storage for the management and auditing of encrypted data file, which can be used in energy domain. 
Such a system not only enhances data integrity, provenance, and long-term credibility in open science, but also enables reproducible and collaborative workflows while maintaining compliance with privacy regulations such as the GDPR.

 The rest of the paper is structured as follows: Section~\ref{related_work} presents related work. The proposed design of EDGChain-E is introduced in Section~\ref{design}. EDGChain-E is further discussed in Section~\ref{discussion}. Finally,~Section~\ref{conclusion_future_work} concludes the paper. 
 
\section{Related Work}
\label{related_work}
Recent research has explored how blockchain technology can enhance data provenance, integrity, and reproducibility in scientific workflows. Hernandez (2024) presented  DGChain~\cite{hernandez2024dgchain}, a decentralized framework that ensures the integrity, traceability, and transparency of research data throughout its lifecycle, supporting secure and collaborative science in distributed settings.

In parallel, several systems have aimed to address secure data management in decentralized environments. One notable example is DSDOS (Decentralized Secure Data Outsourcing System), a blockchain-based framework designed for secure and verifiable data outsourcing in untrusted cloud infrastructures~\cite{dsdos}. 
DSDOS employs hybrid encryption, combining symmetric encryption such as AES for data confidentiality and asymmetric encryption for key distribution. 
Smart contracts enforce access control, 
while the blockchain ledger stores cryptographic hashes and metadata to ensure integrity and auditability. 
By integrating encryption with decentralized consensus, DSDOS provides trustless, tamper-resistant storage without depending on centralized authorities.

Extending these paradigms, EDGChain-E targets the specific needs of collaborative  data management for energy domain. It integrates versioning capabilities similar to Git, persistent identifier support like Zenodo, and the integrity guarantees of blockchain technology. Unlike DSDOS, which emphasizes secure outsourcing, EDGChain-E focuses on data provenance, branching history, and reproducibility in eScience workflows. Its encryption-aware design allows privacy-preserving dataset sharing while maintaining verifiability and auditability in multi-party research collaborations.

The EDGChain-E framework could complement prior blockchain-based eScience efforts by providing secure and decentralized data versioning, in line with applications such as the autonomous computational broker for scientific computing resources by Alimoğlu and Özturan~\cite{ebb_article}, the smart contract–based autonomous organization for sustainable software~\cite{eScience/AlimogluO17}, and the blockchain-based workflow execution broker for eScience workflows~\cite{alimoglu_autonomous_2024}.

Finally, EDGChain-E encrypts data to protect privacy, but it still follows
the principles of the Open Science Framework (OSF)~\cite{foster2017open}. OSF supports
clear data standards, transparent data history tracking, and easy data
sharing. EDGChain-E supports these goals by providing transparent and
traceable data processes that protect both privacy and research integrity.

Table~\ref{tab:framework-comparison} presents a comparative analysis of the widely used frameworks alongside the proposed approach, highlighting their respective strengths and limitations.

\section{Design of EDGChain-E}
\label{design}

\begin{figure}[t]
    \centering
     \vspace{0.5em}
    \includegraphics[width=0.4\textwidth]{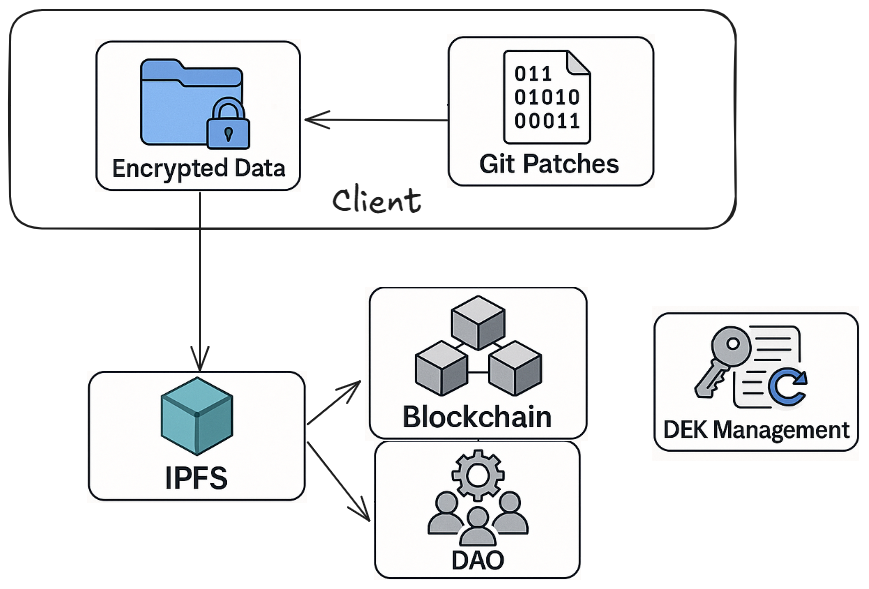}
        \caption{Components of EDGChain-E.}
    \label{fig:components}
\end{figure}

The EDGChain-E system consists of three primary layers: the client interface, the Ethereum smart contracts\footnote{The smart contract of EDGChain-E is available within the following GitHub repository: https://github.com/avatar-lavventura/EDGChain-E}, and the IPFS network. Figure~\ref{fig:components} shows the components of EDGChain-E. 

\begin{figure}[t]
    \centering
     \vspace{0.6em}
    \includegraphics[width=0.48\textwidth]{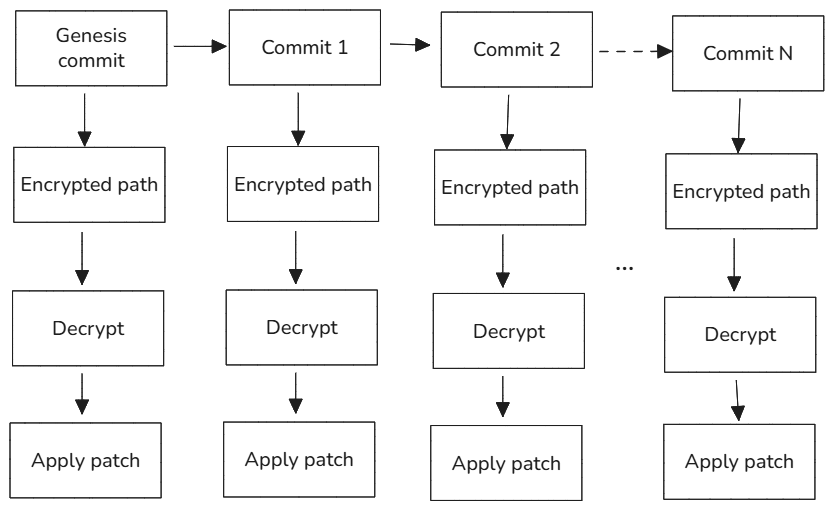}
        \caption{Reconstruction of the full version history timeline.}
    \label{fig:commit_n}
\end{figure}

The data commit workflow represents a critical stage in our implementation. 
The typical process for a data commit is outlined as follows:
\begin{enumerate}
    \item \textbf{Local Data Creation}: A user or a group of users creates the genesis file if the file is initial version, otherwise they create patch files which represent the changes with respect to the genesis file.
    \item \textbf{Hybrid Encryption}: The client generates a fresh Data Encryption Key (DEK), typically a 256-bit key derived from a cryptographically secure pseudorandom number generator (CSPRNG).
    The data is encrypted using AES-256 in GCM mode with the given DEK, 
    using the function 
    \verb|EncryptedData=Encrypt(PlainData, DEK)|. 
    The DEK is encrypted once for the entire authorized group using a group encryption scheme (e.g., broadcast encryption or group key encapsulation), enabling efficient and simultaneous key distribution while preserving confidentiality. 
      
    \item \textbf{IPFS Storage}: 
    The encrypted genesis file and its patches are uploaded to IPFS, typically via its HTTP API.    
    IPFS returns CIDs that uniquely reference the encrypted objects and ensure integrity through their SHA-256-based construction.
    Note that all uploaded patches up to the latest checkpoint of the genesis file are pinned in IPFS, 
    which prevents them from being removed. 
    The CIDs are obtained from the smart contract, whereas the corresponding encrypted DEKs are distributed by the project owner to members of the collaboration group via email or another secure communication channel.
    \item \textbf{IPFS Pinning}: Since all patches are necessary for data reconstruction, an IPFS node must protect the relevant content from garbage collection by employing the direct pinning strategy. 
    To ensure redundancy and long-term availability, an IPFS cluster can be deployed to store all patches as part of a backup solution.
    \item \textbf{Blockchain Commit}: The client submits a transaction to the EDGChain-E smart contract, using a tool such as \texttt{web3.py}. The transaction includes parameters such as the new CID, parent CID, and encrypted key references (e.g., via a function like \texttt{commitData(newCid, parentCid, encryptedKeys)}), and is cryptographically signed by the user's Ethereum private key. This step records the new version immutably on-chain.    
    \item \textbf{On-Chain State Update}: Upon confirmation, the smart contract registers the new commit by linking the new CID to its parent and storing the encrypted key metadata. An event is emitted to notify any network listeners of the update.    
    \item \textbf{Secure Retrieval}: Users can query the smart contract to obtain the latest CID, while only authorized users are able to decrypt their corresponding encrypted DEK.    
    The client then downloads the encrypted file from IPFS, decrypts the DEK using the user's private key (e.g., via GPG or Ethereum-based decryption), and decrypts the file content to recover the plaintext.
    \item \textbf{Full History Reconstruction}: 
    To fully reconstruct the dataset as of commit~$n$, 
    the EDGChain-E client performs a recursive retrieval and patch application process. This is outlined next step by step. \textbf{Fetch Commit History}: Starting from the latest commit~$n$, the client queries the smart contract to traverse backwards through the chain of parent CIDs until it reaches the genesis commit.  \textbf{Download Encrypted Blobs}: For each commit in the lineage $[0, \ldots, n]$, the client retrieves the encrypted data file or patch blob from IPFS using its CID.
    \textbf{Decrypt and Apply Patches}: For each retrieved blob: The client decrypts the associated DEK using the user’s private key (e.g., via GPG or ECDH-based decryption); The encrypted data is then decrypted with the DEK using AES-256-GCM, and the resulting plaintext (either a full file or a patch) is sequentially applied to reconstruct the state at commit~$n$. 
\textbf{Final State Reconstruction}: After applying all patches in order, the client obtains the final, complete dataset at commit~$n$. 

This enables any authorized participant to reconstruct the full dataset history with integrity and confidentiality preserved.
Figure~\ref{fig:commit_n} illustrates the reconstruction of the complete version history.
\end{enumerate}

This design ensures that all cryptographic operations, including encryption, key generation, and decryption, are confined to the client side.
The blockchain layer stores only metadata and encrypted references. 
All actions, including commits, key grants, and versioning, are cryptographically signed and timestamped, offering a transparent and verifiable audit trail.
In this approach, IPFS stores only encrypted data file and therefore does not have access to the plaintext content at any stage. This design guarantees end to end verifiability and confidentiality, ensuring that only authorized users with the correct decryption keys can reconstruct the original file, while intermediaries including IPFS remain unable to access the plaintext data file.

\section{Discussion on EDGChain-E}~\label{discussion}
The on-chain layer of EDGChain-E is built around Solidity smart contracts that manage references to encrypted data file.
Each commit record in the smart contract includes the IPFS CID of the encrypted payload (either a full file or a patch), 
a pointer, which is an IPFS hash, to its parent commit, 
Role-based permissions (e.g. owner, contributor, reviewer) are enforced via modifieras: 
only authorized callers can publish new commits or modify access lists.


The version history is maintained as a linked list of commits.
Each new commit includes a reference to its parent, forming a cryptographically verifiable sequence of revisions. Because IPFS CIDs are content-addressed (using SHA-256 hashes), any alteration of stored data file is detectable.  
The smart contract emits an event for each commit, so off-chain agents can track the latest CID and the topology of changes.

After a predefined number of commits (N), a new genesis state can be established. This mechanism prevents the unbounded accumulation of patches and limits the depth of the commit history that must be applied for reconstruction.
Version tracking is achieved by linking each commit record to its parent, thereby forming an immutable revision chain. The CID of each encrypted object is logged on-chain, enabling clients to retrieve any version from IPFS and verify its integrity using SHA-256 hashing.
Note that a single DEK is used for each genesis file and all of its patches until the next genesis file.



In IPFS, each file is identified by a unique cryptographic hash (CID), which makes it immutable and enables efficient versioning. In the context of versioning encrypted data files, IPFS facilitates secure, decentralized storage of encrypted files, with each new version generating a unique CID. This mechanism guarantees data integrity, as any modifications to the file result in a change to its hash. As a result, encrypted files can be uploaded to IPFS, and any changes will automatically trigger the creation of a new CID, enabling clear version control without the need for centralized systems. 
Files can be retrieved from any node storing the content, ensuring redundancy and availability. 
IPFS can also be seamlessly integrated with blockchain and smart contracts to automate data provenance, enforce access rules, and verify file integrity, making it a comprehensive solution for secure data management and versioning in research and other applications.


In this decentralized framework, IPFS plays a pivotal role not only in storing but also in managing file transfers. 
Since IPFS operates in a peer-to-peer manner, files are transferred directly between users, avoiding centralized intermediaries. This decentralized nature means that, when a requested file object is unavailable on a node, the file will be fetched from another node within the network, ensuring that users can still access the content even if it is not locally stored. This unique approach to file retrieval underscores the flexibility and resilience of IPFS as a decentralized storage system, where data is not bound to a single location or authority.


In our approach, within decentralized systems like IPFS, a hybrid encryption scheme is employed: the sender encrypts plaintext data with a symmetric Data Encryption Key (DEK) (e.g., AES) for efficiency, then encrypts the DEK separately for each recipient using their public keys (e.g., RSA or GPG), creating a single DEK file containing encrypted segments for all authorized users. When new users are added, the DEK file is updated and stored as a new version.

Authorized users obtain the encrypted DEK file through secure channels such as email, decrypt their respective DEK segments, and subsequently decrypt the data using the recovered DEK.
This approach streamlines key management by requiring updates only to the DEK file upon adding recipients. By consolidating multiple recipients’ keys into one DEK file, the method reduces computational overhead while preserving privacy and scalability. The hybrid system—symmetric encryption for data and asymmetric encryption for key distribution—offers an efficient, privacy-preserving solution for decentralized data sharing.




\begin{figure}[t]
    \centering
     \vspace{0.5em}
    \includegraphics[width=0.5\textwidth]{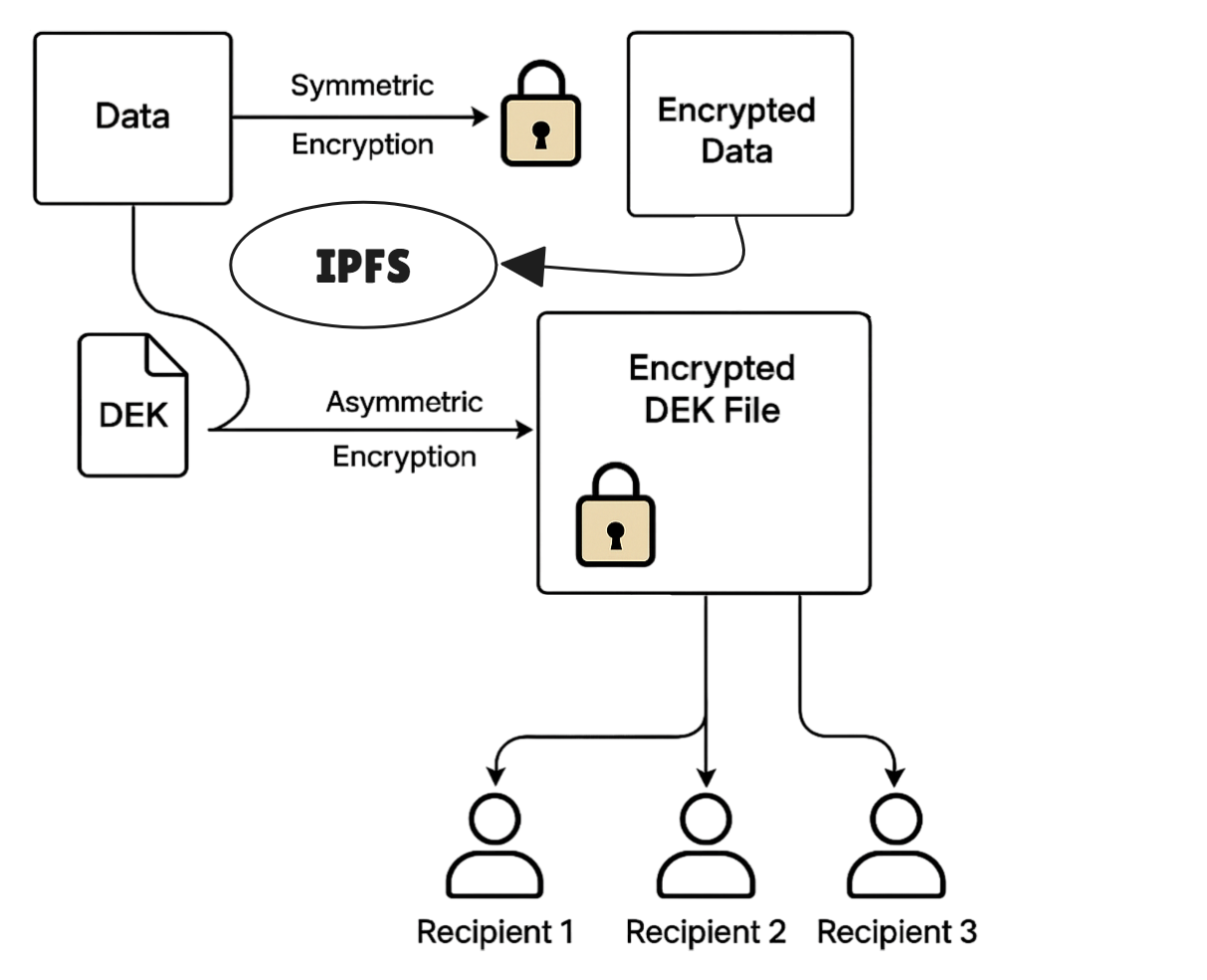}
        \caption{Secure Data Sharing in Decentralized Systems Using Symmetric Encryption and DEK File Management.}
    \label{fig:dek}
\end{figure}

Figure~\ref{fig:dek} depicts the architecture. The project owner shares DEK with the collaborators by email or other secure communication channels. The  data itself is symmetrically encrypted using the DEK and stored on IPFS, ensuring decentralised and tamper-resistant storage. To enable secure access by the collaborators, the DEK is encrypted separately with each recipient’s public key, forming the DEK file. Collaborators can retrieve the encrypted data file from IPFS after decrypting the DEK using their own private keys to access the original content. Unauthorized users, lacking the appropriate private key, are unable to recover the DEK, thereby maintaining confidentiality. 



To optimize storage and minimize computational overhead, the system transmits only the differences between versions rather than re-encrypting the entire file. 
This approach reduces storage requirements and improves efficiency by ensuring that only the modified portions of the data are updated.

\noindent This incremental design of DGChain-E ensures:
\begin{itemize}
    \item \textbf{Confidentiality}: All patches are encrypted before being stored in IPFS; only authorized users with the correct keys can decrypt them and reconstruct the versions.
    \item \textbf{Storage Efficiency}:  Small sized patches are stored, reducing IPFS storage usage and minimizing data transferred on updates.
    \item \textbf{Integrity}: IPFS content addressing via SHA-256 CIDs ensures that any tampering is immediately detectable.
    \item \textbf{Auditability}: The blockchain stores a verifiable, immutable log of all version updates and their relationships.
\end{itemize}

\textit{{Cryptographic Primitives used for DGChain-E:}}
EDGChain-E integrates standard cryptographic primitives to ensure confidentiality, integrity, and authenticity:
\begin{itemize}
    \item \textbf{Symmetric Encryption (AES-256)}: Data and patches are encrypted using AES-256, typically in authenticated modes like GCM, which provide both confidentiality and integrity via authentication tags. If used in unauthenticated modes (e.g., CBC, CTR), integrity is ensured separately using HMAC-SHA256 or Merkle hashing.
    \item \textbf{Asymmetric Encryption (RSA/ECC)}: DEK is encrypted per recipient using public-key cryptography, such as RSA-2048 (GPG) or ECIES with Ethereum-compatible secp256k1 keys. Only holders of the corresponding private keys can recover the corresponding DEK.
    \item \textbf{Hashing (SHA-256)}: SHA-256 is employed for content addressing in IPFS, ensuring immutable and verifiable object integrity. It is also used for commitments and checksums throughout the protocol.
    \item \textbf{Digital Signatures}: All blockchain transactions are signed with ECDSA via users' Ethereum accounts, binding actions to identities. Optionally, GPG signatures are used for key verification.
    \item \textbf{Randomness}: DEK, nonces, and salts are generated using cryptographically secure pseudo-random number generators to ensure unpredictability.
\end{itemize}

By combining authenticated encryption with public-key wrapping, EDGChain-E ensures end-to-end security: encrypted data file is tamper-evident, key distribution is cryptographically protected, and blockchain guarantees transparency and immutability.


\subsection{Security Considerations}
EDGChain-E's security model rests on client-side cryptography, immutable logs, and decentralized governance. 
Key security properties include:
\begin{itemize}
    \item \textbf{Confidentiality}: Achieved through end-to-end encryption.
    \item \textbf{Integrity}: Provided by IPFS's SHA-256 content addressing.
    \item \textbf{Auditability}: All commits and policy changes are logged on-chain.
    \item \textbf{Non-repudiation}: All transactions are signed and timestamped on the blockchain.
\end{itemize}

\section{Conclusion and Future Work}
\label{conclusion_future_work}
EDGChain-E is proposed as a decentralized, encrypted data versioning platform specifically designed for Energy domain. The system incorporates a hybrid encryption scheme, which we have demonstrated to provide a robust, efficient, and privacy-preserving mechanism for secure data sharing in decentralized infrastructures. By carefully balancing strong security guarantees with features that facilitate practical collaboration, EDGChain-E aims to support workflows involving sensitive or high-value Energy datasets.

This framework seeks to address key limitations observed in existing systems. EDGChain-E integrates the versioning capabilities of Git, the immutability and cryptographic assurance of blockchain technologies into a unified platform. In scenarios involving sensitive or encrypted data, EDGChain-E ensures confidentiality through end-to-end encryption and data integrity through on-chain cryptographic hashing—capabilities that are not simultaneously offered by centralized repositories or conventional version control systems for energy domain. As a result, EDGChain-E enables privacy-preserving data sharing among research collaborators, establishing a secure and transparent foundation for decentralized collaboration in energy domain.

As future work, the smart contract could be extended to support grant access functionality, such as specifying read and write permissions for individual users.
In addition, new genesis files could be created for each branch, with patches built incrementally on top of them to enable parallel version histories.
Furthermore, Python-based scripts could be developed to facilitate user interaction with the smart contract and streamline workflow.

\bibliographystyle{IEEEtran}
\bibliography{bibl}

\end{document}